\documentclass[fleqn,twoside]{article}
\usepackage{espcrc2}

\usepackage[dvips]{graphicx}

\title{
\vspace{-1.8cm}
\hfill \rm \null \hfill
\hbox{\normalsize ADP-02-89/T528} \\
\vspace{+1.3cm}
FLIC Overlap Fermions}

\newcommand{\eqn}[1]{ \begin{equation} #1 \end{equation} }

\author{W. Kamleh\address[CSSM]{Special Research Centre for the Subatomic Structure of Matter (CSSM) and Department of Physics and Mathematical Physics, University of Adelaide 5005, Australia.},
D.B. Leinweber\addressmark[CSSM],
A.G. Williams\addressmark[CSSM],
J.B. Zhang\addressmark[CSSM]}

\begin{document}

\thispagestyle{empty}

\begin{abstract}
The action of the overlap-Dirac operator on a vector is typically
implemented indirectly through a multi-shift conjugate gradient
solver.  The compute-time required depends upon the condition number,
$\kappa$, of the matrix that is used as the overlap kernel.  While the
Wilson action is typically used as the overlap kernel, the FLIC (Fat
Link Irrelevant Clover) action has an improved condition number and
provides up to a factor of two speedup in evaluating the overlap
action.  We summarize recent progress on the use of FLIC overlap
fermions.
\end{abstract}

\maketitle

\section{Introduction}

Overlap fermions \cite{overlap4} are a realisation of chiral symmetry
on the lattice. Given some reasonable Hermitian-Dirac operator $H$, we
can deform $H$ into a chiral action through the overlap formalism,
\eqn{D_o = \frac{1}{2}\big( 1+\gamma_5 \, \epsilon(H) \big), \quad
\epsilon(H)=\frac{H}{\sqrt{H^2}}. } 
Unfortunately, the matrix sign function $\epsilon(H)$ is difficult to
evaluate and is typically approximated by a sum over poles
\cite{neuberger-practical} which can be evaluated using a multi-shift
conjugate gradient (CG) solver \cite{edwards-practical}. This is an
iterative approximation where the number of iterations for a given
accuracy increases with the condition number of the kernel, $\kappa(H)
= |\lambda_{\rm max}/\lambda_{\rm min}|$.

Usually the Hermitian Wilson-Dirac operator is used as the overlap
kernel. Its low-lying spectrum is characterised by a handful of
isolated eigenmodes which can be very small, increasing the condition
number, $\kappa$, unacceptably. These eigenmodes can be projected out
of the basic operator, reducing its condition number to a numerically
acceptable level, and then dealt with explicitly
\cite{edwards-chiral}. Unfortunately, as the spectrum rapidly becomes
dense, projecting out low-lying modes can only help one so far.

An alternative is to use a kernel with an improved spectrum, that is,
where the region of dense modes is shifted away from zero. The FLIC
(Fat Link Irrelevant Clover) \cite{zanotti-hadron,kamleh-overlap}
action possesses this property.

\section{Spectral Flow and Condition Number}

\begin{figure*}[!tb]
\includegraphics[height=0.48\textwidth,angle=90]{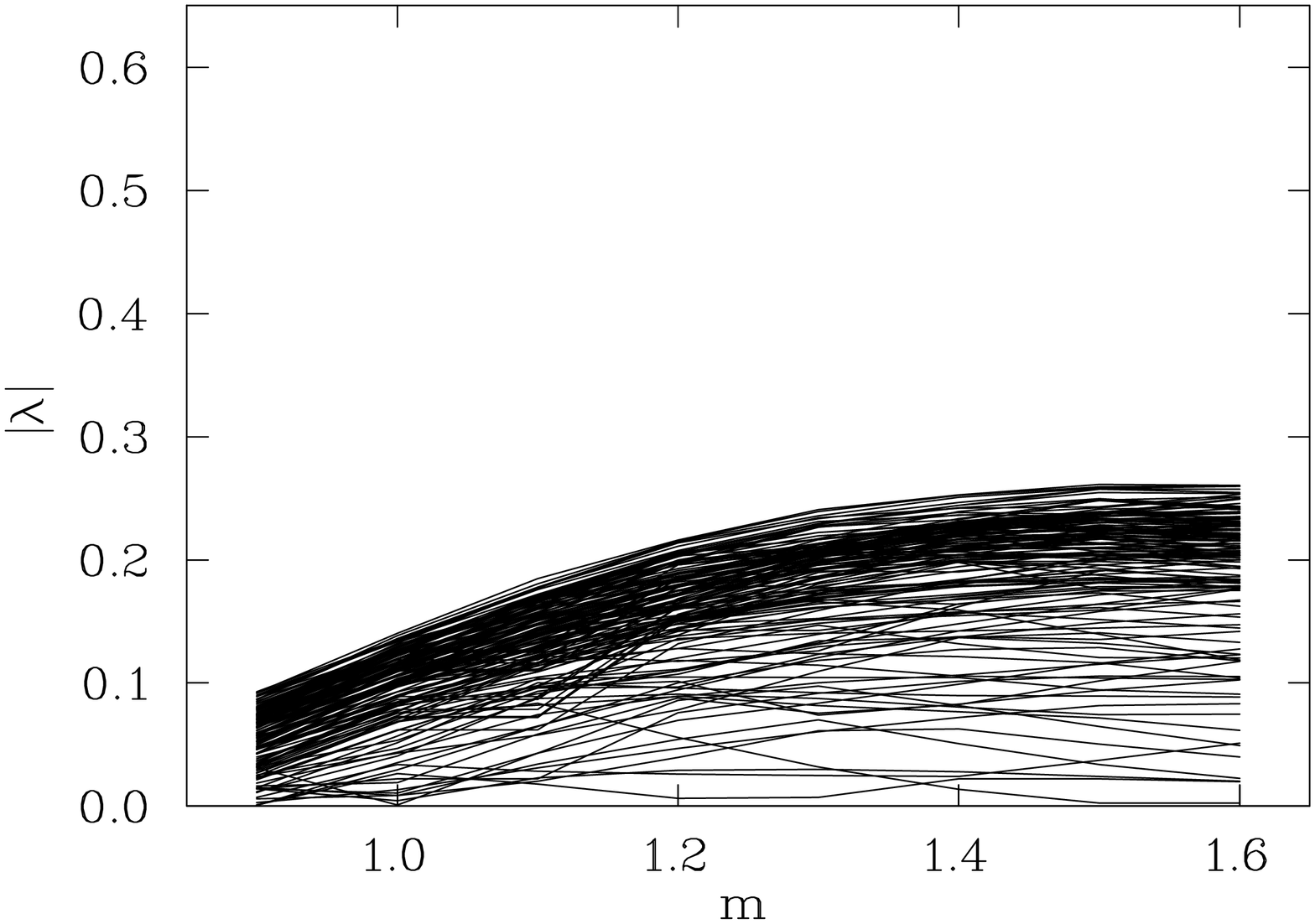}
\includegraphics[height=0.48\textwidth,angle=90]{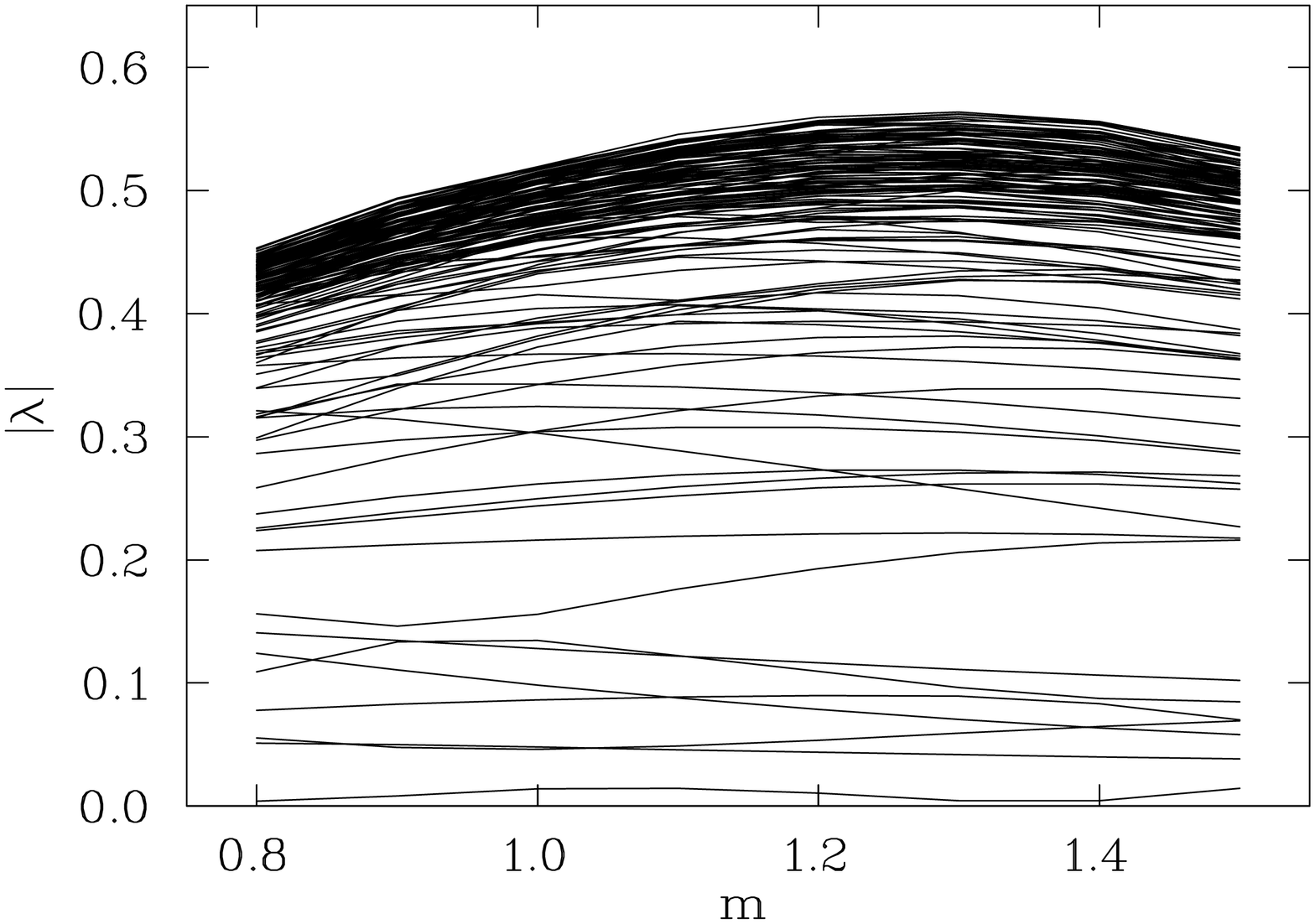}
\vspace{-36pt}
\caption{Spectral flow of the Wilson action (left) and the FLIC action
(right) at $\beta=4.60$. \vspace{-12pt} }
\label{fig:flow}
\end{figure*}

Spectral flow diagrams give us a good comparison of the two different
kernels, FLIC and Wilson, as they allow us to see directly the
difference in the qualitative structure of the low-lying spectra of
the two actions \cite{kamleh-overlap}.   Figure \ref{fig:flow} displays
the flow of the lowest 15 eigenvalues as a function of $m$ for an
ensemble of 10 mean-field improved Symanzik configurations at $\beta =
4.60$ and size $12^3\times24$, with $a=0.122(2)$. As we are interested
in the magnitude of the low-lying values rather than their sign, we
illustrate $|\lambda|$ vs $m$.

\begin{figure*}[!tb]
\includegraphics[height=0.48\textwidth,angle=90]{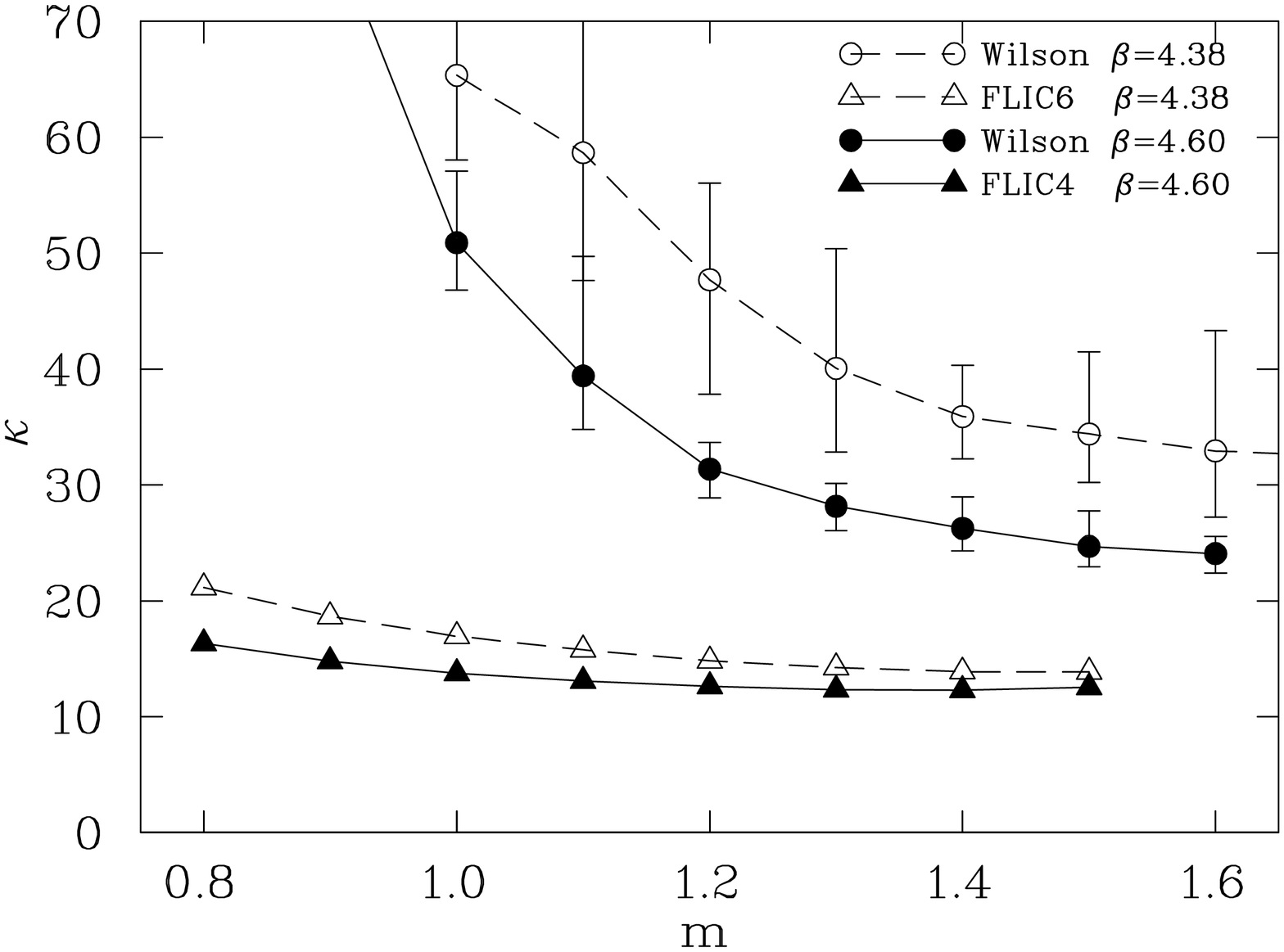}
\includegraphics[height=0.48\textwidth,angle=90]{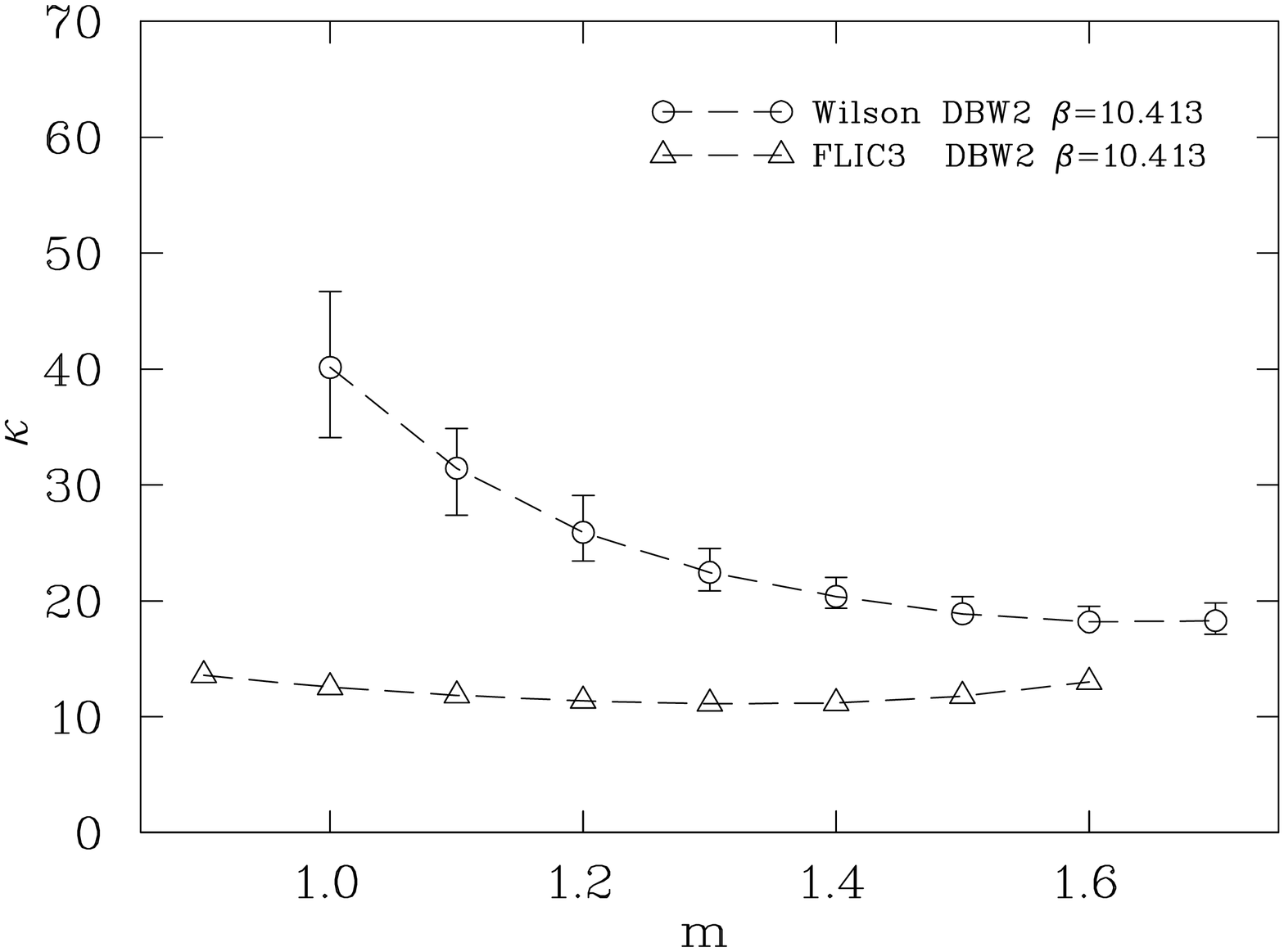}
\vspace{-24pt}
\caption{Comparison of the condition number of the FLIC and Wilson fermion actions.
Symanzik improved glue (left plot) at both $a=0.165$ and $a=0.122$ with 15 modes projected out, and
DBW2 glue (right plot) at $a=0.165$ with 10 modes projected out are illustrated.}
\label{fig:kappa}
\end{figure*}

We see the Wilson spectrum is very poor, with a high density of very
small eigenmodes and no gap away from zero.  By contrast, the FLIC
action (which has a clover term, and irrelevant operators constructed
from four-sweep APE-smeared links) possesses a spectrum which is
clearly superior to that of the Wilson. Not only has the density of
very small modes been significantly reduced, the region where the
spectrum becomes dense has been shifted much further away from zero.

This improvement in the spectrum results in the FLIC action being much
better conditioned than the Wilson action. In Figure \ref{fig:kappa}
we examine the condition number, $\kappa$, of the two actions as a
function of $m$. The condition number is calculated after
having projected out the lowest 15 eigenmodes on the 2 lattices that
are shown. The finer lattice is the same as the one used in the
spectral flow plots, and the coarser lattice uses the same gluonic
action, but is an $8^3\times16$ lattice at $\beta=4.38$, corresponding
to a lattice spacing of a=0.165(2). The points are the mean condition
numbers across the ensembles, and the error bars indicate the minimum
and maximum condition numbers, giving an idea of the variation in
$\kappa$.

\section{Gluonic Action}

There have been suggestions to accelerate the computation of the sign
function by using non-pertubatively improved gauge actions
\cite{edwards-topcharge}.  Our results are based on Symanzik improved
gauge configurations and further improvements arising from the use of
the FLIC action are in addition to that of using improved gluon-field
configurations. This is verified by performing a similar analysis on
gluonic configurations using a Monte-Carlo Renormalisation Group
improved action. Some preliminary results of this investigation using
DBW2 glue \cite{borici-dbw2} are displayed in Figure \ref{fig:kappa},
with a full report given elsewhere \cite{kamleh-dbw2}. We note that at a
fixed lattice spacing, DBW2 glue improves the condition number for 
both actions, although the effect is much more pronounced for the Wilson action.

\section{Compute Time}

Saving iterations (by reducing the condition number) does not
necessarily reduce the most important quantity, compute time.
Shifting from a standard Wilson action to a partially smeared action
means that we now have two sets of gauge fields, the standard and
smeared links.  Additionally, the standard spin-projection trick is no
longer applicable, possibly providing an additional factor of two in
compute time needed.  However, it can be shown that the spin
projection trick can be generalised to include partially smeared
actions as well \cite{kamleh-spin}.  This results in paying at most a
single factor of two compute-time for a FLIC-fermion matrix-vector
multiplication.  As there is significant additional expense in the
evaluation of the overlap sign function, we get to keep the majority
of the speedup gained by reducing the number of iterations
\cite{kamleh-overlap}. This results in FLIC-Overlap fermions being
approximately twice as fast as the standard Wilson-kernel formulation.

\section{Physical Results}

\begin{figure}[!tb]
\includegraphics[height=0.48\textwidth, width=0.28\textheight, angle=90 ]{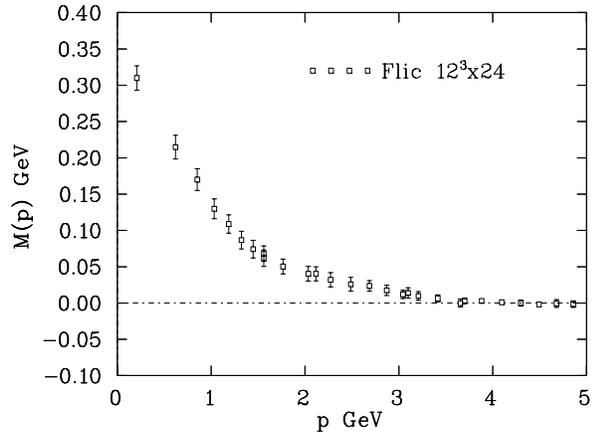}
\vspace{-36pt}
\caption{The chirally extrapolated mass function as a function of $p$,
at $\beta=4.60, L=12^3\times24$.\vspace{-12pt}}
\label{fig:quarkprop}
\end{figure}

Regardless of the kernel used, all overlap fermions are free from
$O(a)$ errors. However, different kernels may in general produce
actions which differ at $O(a^2)$.  As a first investigation into this
matter, we have calculated the quark propagator in momentum space
using FLIC-Overlap fermions, essentially performing the same
calculation that has been done earlier with the standard overlap
action \cite{zhang-overprop}.  The chirally extrapolated mass function
is shown in Figure \ref{fig:quarkprop}, with a full report given
elsewhere \cite{kamleh-flicoverprop}.

\section{Conclusion}

In the overlap formalism one is free to choose the argument of the
sign function, the overlap kernel, so long as one uses a reasonable
Dirac operator.  The standard kernel choice is the Wilson action.  By
choosing the FLIC action, one can obtain a significant reduction in the
compute time needed for overlap fermions.  We have reviewed the work
that has been done to date using FLIC Overlap fermions, including some
preliminary results into the physical structure of the quark
propagator.







\end{document}